# How to Collaborate between Threshold Schemes

Daoshun Wang[1,*], Ziwei Ye[1], Xiaobo Li[2]

[1]Department of Computer Science and Technology, Tsinghua University, Beijing, China

[2]Department of Computing Science, University of Alberta, Edmonton, Alberta, Canada

**Abstract:** Threshold schemes have been used to protect secrets by distributing shares to participants. To protect two secrets, we can use two separate traditional schemes, say, a $(t_1, n_1)$ scheme and a $(t_2, n_2)$ scheme. If there are $u$ ($\leq min(t_1, t_2)$) participants involved in both schemes, each of these $u$ participants must keep two different shares. This paper proposes a method that allows each common participant to keep only one share. Our method constructs two polynomials with $u$ common crossover points. We give theoretical details and two demonstrative examples. This algorithm can also handle the collaboration between more than two schemes.

## 1   Introduction

To protect a secret, such as an encryption key, *threshold scheme* (also called *secret sharing*) has been proposed [3, 6, 5] to divide the secret into *shares* and the shares can be distributed to *participants*. When two or more secrets need to be protected, several separate schemes can be employed. If a participant is involved in two or more schemes, he must keep multiple shares. Consider a case with a (3, 5) scheme and a (4, 6) scheme, and there are two common participants. This is similar to the situation where two banks want to collaborate. Each bank has a safe cabinet. The first bank has a group of five VPs responsible for its cabinet. Any

---

* Corresponding author. Tel.:+86-10-62782930.

*E-mail address:* daoshun@mail.tsinghua.edu.cn  (Daoshun Wang).

three of the five can open it. The second bank has a group of six VPs handling their cabinet. Any four of the six can open it. These two VP groups overlap by two. In other words, two VPs are working for both banks and are involved with both cabinets. This phenomenon has become more commonplace in today's co-operation practice between organizations. With the traditional threshold schemes, each common VP must carry two keys. In the digital case, each common participant must keep two shares, along with their corresponding indices. If a participant is involved with a large number of secrets, he needs to keep many shares which can be a burden. Note that the dealers of the two threshold schemes may not be the same person. One dealer may not know the secret of the other dealer. Therefore, we basically still have two separate schemes but they are to be designed in a way that facilitates collaboration in terms of common participants.

In this paper, we propose a method for constructing two (or more) threshold schemes allowing each common participant to keep only one share. Suppose that we have a $(t_1, n_1)$ scheme and a $(t_2, n_2)$ scheme with $u \leq t_1 \leq t_2$ ( not $u = t_1 = t_2$ ) common participants. Using Shamir's polynomial approach [6], the first curve is decided arbitrarily as long as $p$ is chosen with both schemes in mind. From the common crossover points, the second curve can be easily constructed.

The detailed construction method for a simple two-scheme case is presented in Section 2. Section 3 discusses cases with more than two schemes. Section 4 gives conclusions. Two numerical examples are shown in Appendix.

## 2  A Simple Case with Two Secrets

### 2.1 Curve Construction

To protect two secrets ( $a_{1,0}$ and $a_{2,0}$ ) using polynomial interpolation [1] proposed by Shamir [6], a $(t_1, n_1)$ scheme and a $(t_2, n_2)$ scheme are used, where $a_{1,0}, a_{2,0}, n_1, n_2 \in Z^+$ . Without loss of generality, let $t_1 \leq t_2$ and suppose there are $u \leq t_1$ common participants. If $u = t_1 = t_2$ and the two secrets are different numbers, the traditional threshold schemes must be used and each common participant must keep two shares. If the three values ( $u$ , $t_1$ and $t_2$ ) are not all equal, i.e., either $u < t_1$ or $t_1 < t_2$, the proposed construction algorithm below can guarantee that each common participant keeps only one share.

To begin with, the dealers agree on a large prime number $p$ with $p \geq \max(a_{1,0}, a_{2,0}, n_1+1, n_2+1)$. Let $Z_p$ represent the field of integers modulo $p$. Choose $a_{1,i_1} \in Z_p$, $a_{2,i_2} \in Z_p$, where $i_1 = 0, \cdots t_1-1$, $i_2 = 0, \cdots, t_2-1$. The first polynomial is

$$f_1(x) = a_{1,0} + a_{1,1}x + a_{1,2}x^2 + a_{1,t_1-1}x^{t_1-1} \pmod{p} \tag{1}$$

Let $D_{1,i} = f_1(i)$, where $1 \leq i \leq n_1$, denote the evaluation of $f_1(i)$ over $Z_p$. Dealer 1 chooses any $u$ points from the $n_1$ points $(1, D_{1,1}), \cdots, (n_1, D_{1,n1})$ in the 2D plane. Suppose that he picked $(1, f_1(1)), (2, f_1(2)), \cdots, (u, f_1(u))$ and passed on to Dealer 2.

Then Dealer 2 constructs $f_2(x)$ starting from $(0, a_{2,0})$. He randomly picks $t_2 - u - 1$ points $(u+1, f_2(u+1)), \cdots, (t_2-1, f_2(t_2-1))$. One way to do it is to compute any arbitrary polynomial of degree $t_2 - 1$.

Some special cases may need attention. When $t_1 = t_2$, the polynomial curves $f_1(x)$ and $f_2(x)$ are of the same degree. As long as the constant terms (the secrets they are protecting) are different, i.e., $a_{1,0} \neq a_{2,0}$, Dealer 2 can proceed as usual. Example 2 demonstrates this situation in Appendix. When $t_1 = t_2$ and $a_{1,0} = a_{2,0}$, Dealer 2 should avoid a "conflict", i.e., curves $f_1(x)$ and $f_2(x)$ could end up to be identical to each other.

Since each coefficient can take $p$ possible values, the probability that $f_2(x)$ is identical to $f_1(x)$ is $1/(p^{t_2-u-1})$ which is very close to zero. In other words, the set $\{(u+1, f_2(u+1)), \cdots, (t_2-1, f_2(t_2-1))\}$ that Dealer 2 selected is almost certainly different from the set of Dealer 1: $\{(u+1, D_{1,u+1}), \cdots, (n_1, D_{1,n1})\}$. At least one point would be different, i.e., $(i, f_2(i)) \neq (i, f_1(i))$. Therefore, $f_2(x)$ being identical to $f_1(x)$ is highly unlikely. To completely avoid this extremely rare conflict, Dealer 2 could send a subset of his points to Dealer 1. If this subset happens to be the same as Dealer 1's points, Dealer 1 can warn Dealer 2 about a possible conflict. He doesn't need to know the entire set of Dealer 2. In application, choosing a big enough value for $p$ would solve this problem quite sufficiently.

So far, Dealer 2 has $t_2$ points:

$(0, a_{2,0})$, $(1, f_1(1)), \cdots, (u, f_1(u))$, $(u+1, f_2(u+1)), \cdots, (t_2-1, f_2(t_2-1))$. For convenience, let's call these points $(i, \varphi(i))$ $(i = 0, \cdots t_2 - 1)$ and apply them to the interpolation polynomial in the Lagrange form

$$f_2(x) = \sum_{j=0}^{t_2-1} \varphi(x_j) \prod_{\substack{l=1 \\ l \neq j}}^{t_2-1} \frac{(x - x_l)}{(x_j - x_l)} \pmod{p}. \tag{2}$$

The result polynomial $f_2(x)$ of degree $t_2 - 1$ will take the following form:

$$f_2(x) = a_{2,0} + a_{2,1}x + a_{2,2}x^2 + a_{2,t_2-1}x^{t_2-1} \pmod{p} \tag{3}$$

A numerical example is given in Appendix to demonstrate the above procedure.

In Equation (3), $f_2(x)$ is formed by an addition of $t_2$ polynomials of degree $t_2 - 1$, each result coefficient is a product of $t_2$ coefficients, and each coefficient takes value in the range of $\{0, \cdots, p-1\}$. When every value in the range is equally likely, the probability of $a_{2,t_2-1} = 0 \pmod{p}$ is $1/p$. When $a_{2,t_2-1} = 0 \pmod{p}$ happens, Dealer 2 should re-select his points so that $a_{2,t_2-1} \pmod{p} \neq 0 \pmod{p}$.

## 2.2 Share Distribution

Dealer 1 distributes $D_i = f_1(i)$ to participants $i$, where $1 \leq i \leq n_1$. Note that the evaluation of $f_1(i)$ in Equation (1) is done over $Z_p$.

Dealer 2 obtains $n_2$ values $\{D'_j = f_2(j), 1 \leq j \leq n_2\}$ by evaluating $f_2(j)$ over $z_p$. He eliminated $u$ values in $\{D'_j\}$ that are identical to those in $\{D_i\}$, then distribute the remaining to the $n_2 - u$ participants.

## 2.3 Secret Reconstruction：

The reconstruction process (to reveal the secret) is exactly the same as that in Shamir's scheme [6].

Given any subset of $t_1$ of the $n_1$ values $D_1, \cdots, D_{n1}$, we can find the coefficients of $f_1(x)$ by interpolation [1,4], and then evaluate

$$D = f_1(0) = (-1)^{t_1-1} \sum_{j=1}^{t_1} \varphi(x_j) \prod_{\substack{l=1 \\ l \neq j}}^{t_1} \frac{x_l}{(x_j - x_l)} \pmod{p}, \text{ where } x_j, x_l \in (1, 2, \cdots, n_1).$$

to reveal the first secret.

Given any subset of $t_2$ of the $n_2$ values $D'_1, \cdots, D'_{n_2}$, we can find the coefficients of $f_2(x)$ by interpolation ([4]), and then evaluate

$$D' = f_2(0) = (-1)^{t_2-1} \sum_{j=1}^{t_2} \varphi(x_j) \prod_{\substack{l=1 \\ l \neq j}}^{t_2} \frac{x_l}{(x_j - x_l)} \pmod{p}, \text{ where } x_j, x_l \in (1, 2, \cdots, n_2)).$$

to reveal the second secret.

Knowing only $t_1 - 1$ (resp. $t_2 - 1$) values, on the other hand, does not suffice in order to discover $D$ (resp. $D'$).

## 2.4 Parameter Consideration:

The above algorithm can not be used when $u = t_1 = t_2$. If such a situation exists and the two secrets are different, traditional threshold schemes must be employed. Having $u = t_1 = t_2$ is not desirable from the security point of view, since the $u$ common participants, possibly from outside, can access the secret without the knowledge of non-common (pure local) participants. It is safer that more than $u$ participants are always required in the reconstruction of the secret, i.e., $u < t_1$ and $u < t_2$.

## 3 Cases with More Than Two Secrets

When there are $s(s \geq 2)$ secrets to be protected, multiple threshold schemes ($(t_1, n_1)$, $(t_2, n_2)$, ..., $(t_s, n_s)$) can be used. If there are $u$ common participants, we can construct $s$ polynomials $f_1(x)$, $f_2(x), \ldots, f_s(x)$ with $u$ common crossover points. Here, the polynomials are

$$f_1(x) = a_{1,0} + a_{1,1}x + a_{1,2}x^2 + \cdots + a_{1,t_1-1}x^{t_1-1} \pmod{p}$$

$$f_2(x) = a_{2,0} + a_{2,1}x + a_{2,2}x^2 + \cdots + a_{2,t_2-1}x^{t_2-1} \pmod{p}$$

$$f_3(x) = a_{3,0} + a_{3,1}x + a_{3,2}x^2 + \cdots + a_{3,t_3-1}x^{t_3-1} \pmod{p}$$

$$\vdots$$

$$f_s(x) = a_{s,0} + a_{s,1}x + a_{s,2}x^2 + \cdots + a_{s,t_s-1}x^{t_s-1} \pmod{p}$$

## 3.1 Curve Construction

Dealer 1 is the first dealer to act and he uses $f_1(x)$ to protect Secret 1 ($a_{1,0}$). He obtains $n_1$ points $(1, f_1(1)), \cdots, (n_1, f_1(n_1))$ in the 2D plane. He can choose any $u$ points among them. For convenience, he picks the first $u$ points, i.e., $(1, f_1(1)), (2, f_1(2)), \cdots, (u, f_1(u))$ for the common crossover points of all polynomials.

For $j = 2, \cdots, s$, Dealer $j$ uses a polynomial $f_j(x)$ of degree $t_j - 1$ with $a_{j,0}$ being Secret $j$. With $(0, a_{j,0})$ as the original point and $u$ common crossover points, he can construct $f_j(x)$.

For example, when $j = 2$, Dealer 2 can obtain $t_2 - u - 1$ points by computing any polynomial $f_2(x)$ of degree $t_2 - 1$ or randomly chooses $t_2 - u - 1$ points. As long as that not all these $t_2 - u - 1$ points come from $\{(u+1, f_1(u+1)), \cdots, (n_1, f_1(n_1))\}$, i.e., at least there is one $i$ with $(i, f_2(i)) \neq (i, f_1(i)), (u+1) \leq i \leq t_2$, $f_2(x)$ would not be identical to $f_1(i)$. The probability that the constant terms are the same between the two polynomials is 1/p. Since the $t_2 - u - 1$ coefficients are random chosen from $p$ integers, the probability of two polynomials being identical is only $1/(p^{t_2-u})$. As we mentioned in Sec.2,1, choosing a large enough $p$ should avoid this extremely rare conflict.

Dealer $j$ with $2 < j < s$ acts similarly. Dealer s can obtain $t_s - u - 1$ points by computing any polynomial $f_s(x)$ of degree $t_s - 1$ or randomly chooses $t_s - u - 1$ points. Not all these $t_s - u - 1$ points should come from the set $\{(u+1, f_1(u+1)), \cdots, (n_1, f_1(n_1))\}$. Namely, at least there is one $i$ with $(i, f_s(i)) \neq (i, f_1(i))$, $(u+1) \leq i \leq t_s$. As we discussed in Section 2.1, the probability of two polynomials constructed by different dealers end up identical to each other (a conflict) is very small, especially when the $p$ value is sufficiently large. If the dealers decide to completely avoid this highly unlikely conflict, they could send each other a subset of their points and warn each other if the subsets are identical.

In the construction process, if $a_{i,t_j-1}(\mod p) = 0 (\mod p)$, $i = 2,\cdots,s, j = 2,\cdots,s$, we need to change $(u+1, f_i(u+1)),\cdots,(t_i, f_i(t_i))$ points, $i = 2,\cdots,s$ so that $a_{i,t_j-1}(\mod p) \neq 0 (\mod p)$, $i = 2,\cdots,s$, $j = 2,\cdots,s$ in order to make sure that $a_{i,t_j} \neq 0$ $(i = 2,\cdots,s-2, j = u+1,\cdots,t_s-2) (\mod p)$.

## 3.2 Some Discussions

It is obvious that the security of the proposed scheme is the same as the traditional scheme [6]. That is, fewer than $t_s$ of the $n_s$ shares cannot reveal any information about the secret, where $s = 1,\cdots$. The computation complexity of the recovery secret proposed schemes is $O(k\log^2 k)$ [1], which is the same as the one of Shamir's scheme in [6]. For more analysis on threshold scheme see [2, 7].

The proposed algorithm can be modified to handle more complex situations where the collaborations between the schemes are of a different nature. For example, there are five dealers responsible for five secrets with five schemes, and the schemes are of (2, 3), (3, 5), (4, 6), (5, 7) and (7, 9). The (2, 3) scheme and the (3, 5) scheme have 2 common participants. The (4, 6) scheme, the (5, 7) scheme and the (7, 9) scheme have 4 common participants.

The dealers may or may be the same person, and they may or may not know each other's secrets. As long as they pass $u$ shares to the common participants, the curves they construct would be valid. Another point of notice is that the dealers and participants in the proposed schemes, who are similar to those of Shamir's scheme, are honest.

## 4. Conclusions:

This paper proposes an effective collaboration mechanism for two or more threshold schemes to insure that each common participant keeps only one share. They achieve this by constructing polynomials with a number of common crossover points. When the polynomials are sequenced in an increasing order of their degree, the construction is always successful. When the co-operation between organizations becomes more popular and common participants are involved in multiple secrets, the proposed method can reduce the storage requirement and increase the security and convenience.

## Acknowledgments:

This research was supported by the National Natural Science Foundation of China under Grant

61170032.

# Appendix: Two Numerical Examples

**Example 1**: collaboration between a (3, 5) scheme and a (4, 6) scheme.

Suppose there are two secrets (integers 1 and 3) that could be protected by two dealers, separately, with two traditional threshold schemes: one (3, 5) scheme and one (4, 6) scheme. Let's the five participants involved with the first secret form Group 1, and let's the six participants for the second secret form Group 2. These two groups overlap by two. That is, each one of these two common participants must keep two shares. For easy lookup, we list the related parameters here:

$$(t_1, n_1) = (3,5),\ (t_2, n_2) = (4,6),\ u = 2,\ a_{1,0} = 1,\ a_{2,0} = 3.$$

**Part 1 of Example 1: Curve Construction**

*Step 1*: Dealer 1 constructs a (3, 5) threshold scheme for Secret 1 (integer 1).

Using our construction method, the dealers agree on a large prime number $p$ as

$$p = \max(1, 3, 5+1, 6+1) = 7.$$

Then he defines the first curve with $a_{1,1} = 3$ and $a_{1,2} = 2$, that is

$$f_1(x) = a_{1,0} + a_{1,1}x + \cdots + a_{1,t_1-1}x^{t_1-1} \ (\mathrm{mod}\ p) = 1 + 3x + 2x^2 \ (\mathrm{mod}\ p),$$

where $x$=0, 1, 2, … , 6.

When $x$ takes values 0, 1, … , 5, this curve produces six points:

$(0, f_1(0) = 1)$, $(1, f_1(1) = 6)$, $(2, f_1(2) = 1)$, $(3, f_1(3) = 0)$, $(4, f_1(4) = 3)$, $(5, f_1(5) = 3)$.

From the above six points, he picks up two points at random, say, (1, 6) and (2, 1), for the two common participants. These two points are handed over to Dealer 2.

*Step 2*: Dealer 2 constructs a (4, 6) threshold scheme for Secret 2 (integer 3).

With $a_{2,0} = 3$, Dealer 2 already has three points: (0, 3), (1, 6), and (2, 1). To construct a polynomial of degree $t_2 - 1 = 3$, he needs one more point. He chooses (3, y). The value of y can be arbitrary, i.e., he picks the point $((3, f_2(3) = 3)$.

Using Lagrange interpolation[1,4]

$$f_2(x) = \sum_{j=0}^{t_2-1} \varphi(x_j) \prod_{\substack{l=1 \\ l \neq j}}^{t_2-1} \frac{(x - x_l)}{(x_j - x_l)} \pmod{p}$$

Dealer 2 substitutes $x_j$ values with 0, 1, 2 and 3, the $t_2$ addition terms are

$$3 \cdot \frac{(x-1)(x-2)(x-3)}{(0-1)(0-2)(0-3)} = \frac{3(x-1)(x-2)(x-3)}{-6} = 3x^3 + 3x^2 + 5x + 3 \pmod{7},$$

$$6 \cdot \frac{(x-0)(x-2)(x-3)}{(1-0)(1-2)(1-3)} = \frac{6(x-0)(x-2)(x-3)}{2} = 3x^3 + 6x^2 + 4x \pmod{7},$$

$$1 \cdot \frac{(x-0)(x-1)(x-3)}{(2-0)(2-1)(2-3)} = \frac{(x-0)(x-1)(x-3)}{-2} = 3x^3 + 2x^2 + 2x \pmod{7},$$

$$3 \cdot \frac{(x-0)(x-1)(x-2)}{(3-0)(3-1)(3-2)} = \frac{3(x-0)(x-1)(x-2)}{6} = 4x^3 + 2x^2 + x \pmod{7}.$$

Therefore, Dealer 2 obtains

$$f_2(x) = 6x^3 + 6x^2 + 5x + 3 \pmod{7}, \text{ where } x = 0, 1, 2, \ldots, 6.$$

Figure 1 shows the cross points between two curves, $f_1(x)$ and $f_2(x)$, in the 2D space. Because of the use of modular arithmetic, the curves are periodic.

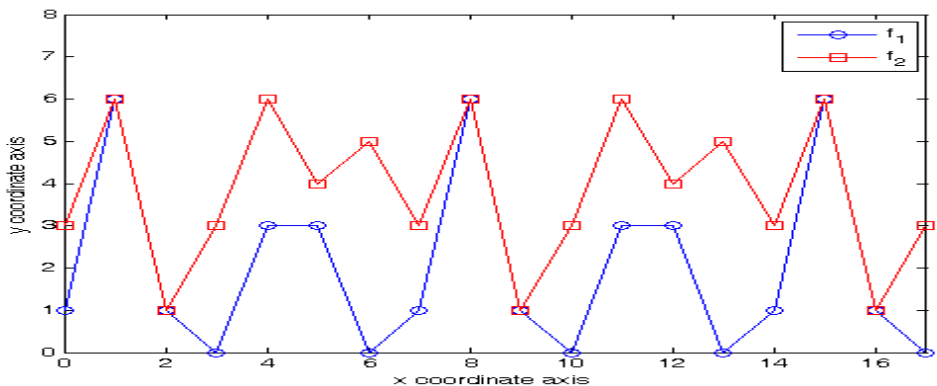

Figure 1. The red curve shows the interpolation polynomial $f_1(x)$ and the blue curve shows $f_2(x)$.

Note that $f_2$ is formed by an addition of $t_2$ polynomials of degree $t_2-1$, each result coefficient is a product of $t_2$ coefficients, and each coefficient takes value in the range of $\{0,...,p-1\}$. When every value in the range is equally likely, the probability of $a_{2,t_2-1} = 0 \pmod{p}$ is $1/p$. In this particular example, this undesirable situation did not occur. The following table lists different result polynomials corresponding to various $y$ values for $(3, f_2(3) = y)$.

Table 1: The effect of the selection of $\varphi(3) \in \{0, \cdots, 7-1\}$ in the form of $f_2(x)$

| $(3, \varphi(3))$ | $f_2(x)$ |
|---|---|
| (3, 0) | $2x^3 + 4x^2 + 4x + 3 \pmod{7}$ |
| (3, 1) | $x^3 + 0x^2 + 2x + 3 \pmod{7}$ |
| (3, 2) | $0x^3 + 3x^2 + 0x + 3 \pmod{7}$ |
| (3, 3) | $6x^3 + 6x^2 + 5x + 3 \pmod{7}$ |
| (3, 4) | $5x^3 + 2x^2 + 3x + 3 \pmod{7}$ |
| (3, 5) | $4x^3 + 5x^2 + x + 3 \pmod{7}$ |
| (3, 6) | $3x^3 + x^2 + 6x + 3 \pmod{7}$ |

It can be seen that each coefficient indeed takes all values of the range with proximately equality frequency, except the constant term. From Table 1, while $\varphi(3) = 2$, dealer 2 obtains $f_2(x) = 0x^3 + 3x^2 + 0x + 3 \pmod{7}$. To avoid a polynomial with highest term being zero, he adds a checking action in the construction phase: when $a_{i,t-1} = 0$, try again and pick a different curve.

**Part 2 of Example 1: Share Distribution**

Using $f_1(x) = (2x^2 + 3x + 1) \mod 7$, $x = 0, 1, 2, \cdots, 5$, Dealer 1 computes $D_i = f_1(i)$ to obtain five points $(1,6), (2,1), (3,0), (4,3), (5,3)$., and distribute to the five participants.

Using $f_2(x) = 6x^3 + 6x^2 + 5x + 3 \pmod 7$, $x = 0, 1, 2, \cdots, 6$, Dealer 2 computes $D'_j = f_2(j)$ to obtain six points $(1,6), (2,1), (3,3),$ and distributes them to the six participants. The first two participants are common between the groups, and each of them needs to keep only one share.

**Part 3 of Example 1: Secret Reconstruction.**

The reconstruction process is exactly the same as that in Shamir's traditional scheme [6].

For Secret 1 protected by the (3, 5) scheme, given any set of $t_1$ of these $n_1$ values ($D_1, \cdots, D_{n_1}$), we can find the coefficients of $f_1(x)$ by interpolation [1, 4], and then evaluate

$$D = f_1(0) = (-1)^{t_1-1} \sum_{j=1}^{t_1} \varphi(x_j) \prod_{\substack{l=1 \\ l \neq j}}^{t_1} \frac{x_l}{(x_j - x_l)} \pmod p, \text{ where } x_j, x_l \in \{1, 2, \cdots, n_1\}. \quad (1)$$

**Case 1**: Both common participants are involved ($(1,6), (2,1)$) plus another participant, one from $\{(3,0), (4,3), (5,3)\}$. Assume that it is $(3,0)$. By (1), we obtain the secret as follows

$$D = f_1(0) = ((-1)^2 \times (6 \times \frac{2 \times 3}{(1-2) \times (1-3)} + 1 \times \frac{1 \times 3}{(2-1) \times (2-3)} + 0 \times \frac{1 \times 2}{(3-1) \times (3-2)})) \pmod 7 = 1$$

**Case 2**: Suppose that one common participant (e.g., (2, 1)) works with two non-common participants (e.g., (4, 3) and (5, 3)). Using Equation (1), we achieve

$$D = f_1(0) = ((-1)^2 \times (1 \times \frac{4 \times 5}{(2-4)(2-5)} + 3 \times \frac{2 \times 5}{(4-2)(4-5)} + 3 \times \frac{2 \times 4}{(5-2) \times (5-4)})) \pmod 7 = 1$$

**Case 3**: Without any common participants, the three non-common participants have three points (3, 0), (4, 3) and (5, 3). We can get

$$D = f_1(0) = ((-1)^2 \times (0 \times \frac{4 \times 5}{(3-4) \times (3-5)} + 3 \times \frac{3 \times 5}{(4-3) \times (4-5)} + 3 \times \frac{3 \times 4}{(5-3) \times (5-4)})) \pmod 7 = 1$$

For Secret 2 protected by the (4, 6) scheme, given any subset of $t_2$ of these $n_1$ values $D'_1, \cdots, D'_{n_2}$, we can find the coefficients of $f_2(x)$ by interpolation [1,4].

**Case 1**: When the two common participants ((1, 6), (2, 1)) and two others (e.g., (4, 6) and (6, 5)) work together, they can obtain the secret by

$$D' = f_2(0) = ((-1)^3 \times (6 \times \frac{2 \times 4 \times 6}{(1-2) \times (1-4) \times (1-6)} +$$

$$1 \times \frac{1 \times 4 \times 6}{(2-1) \times (2-4) \times (2-6)} + 6 \times \frac{1 \times 2 \times 6}{(4-1) \times (4-2) \times (4-6)} + 5 \times \frac{1 \times 2 \times 4}{(6-1) \times (6-2) \times (6-4)})) \pmod 7 = 3$$

**Case 2**: When one common participant (e.g., (1, 6)) works with three non-common participants (e.g., (4, 6), (5, 4) and (6, 5)), they can get

$$D' = f_2(0) = ((-1)^3 \times (6 \times \frac{4 \times 5 \times 6}{(1-4) \times (1-5) \times (1-6)} + 6 \times \frac{1 \times 5 \times 6}{(4-1) \times (4-5) \times (4-6)} +$$

$$4 \times \frac{1 \times 4 \times 6}{(5-1) \times (5-4) \times (5-6)} + 5 \times \frac{1 \times 4 \times 5}{(6-1) \times (6-4) \times (6-5)})) \pmod{7} = 3$$

**Case 3**: Without any common participants, the four non-common participants hold the four points (3,3), (4,6), (5,4) and (6,5). Thus they can compute

$$D' = f_2(0) = ((-1)^3 \times (3 \times \frac{4 \times 5 \times 6}{(3-4) \times (3-5) \times (3-6)} + 6 \times \frac{3 \times 5 \times 6}{(4-3) \times (4-5) \times (4-6)} +$$

$$4 \times \frac{3 \times 4 \times 6}{(5-3) \times (5-4) \times (5-6)} + 5 \times \frac{3 \times 4 \times 5}{(6-3) \times (6-4) \times (6-5)})) \pmod{7} = 3$$

**Example 2:** collaboration between two (3, 5) schemes.

Suppose there are two secrets (integers 1 and 3) that could be protected by two dealers, separately, with two traditional threshold (3, 5) schemes. The first group of five participants involved with the first secret and the second group of five overlap by two. That is, each one of these two common participants must keep two shares. For easy lookup, we list the related parameters here:

$(t_1, n_1) = (3,5)$, $(t_2, n_2) = (3,5)$, $u = 2$, $a_{1,0} = 1$, $a_{2,0} = 3$.

**The curve construction phase for Example 2:**

*Step 1*: (is the same as the Step1 of Example 1)
Dealer 1 defines the first curve $f_1(x) = 1 + 3x + 2x^2 \pmod{7}$, where x=0, 1, 2, ... , 6.
When $x$ takes values 0, 1, ..., 5, this curve produces six points:

$(0, f_1(0) = 1)$, $(1, f_1(1) = 6)$, $(2, f_1(2) = 1)$, $(3, f_1(3) = 0)$, $(4, f_1(4) = 3)$, $(5, f_1(5) = 3)$.

From the above six points, he picks up two points at random, say, (1, 6) and (2, 1), for the two common participants. These two points are handed over to Dealer 2.

*Step 2*: Dealer 2 constructs a (3, 5) threshold scheme for Secret 2 (integer 3).

With $a_{2,0} = 3$, Dealer 2 already has three points: (0, 3), and (1, 6). To construct a polynomial of degree $t_2 - 1 = 2$, he needs one more point..

Using Lagrange interpolation

$$f_2(x) = \sum_{j=0}^{t_2-1} \varphi(x_j) \prod_{\substack{l=1 \\ l \neq j}}^{t_2-1} \frac{(x - x_l)}{(x_j - x_l)} \pmod{p}$$

Dealer 2 substitutes $x_j$ values with 0, 1, and 2, the $t_2$ addition terms are

$$3 \cdot \frac{(x-1)(x-2)}{(0-1)(0-2)} = \frac{3(x-1)(x-2)}{2} = 5x^2 - x + 3 \text{ (mod 7)},$$

$$6 \cdot \frac{(x-0)(x-2)}{(1-0)(1-2)} = \frac{6(x-0)(x-2)}{-1} = -6x^2 + 5x \text{ (mod 7)},$$

$$1 \cdot \frac{(x-0)(x-1)}{(2-0)(2-1)} = \frac{(x-0)(x-1)}{2} = 4x^2 - 4x \text{ (mod 7)},$$

Therefore, Dealer 2 obtains

$$f_2(x) = 3x^2 + 3 \text{ (mod 7)}, \text{ where } x = 0, 1, 2, \ldots, 6.$$

Recall that Dealer 1's curve is

$$f_1(x) = 1 + 3x + 2x^2 \text{ (mod 7)}.$$

These two curves are shown in the following figure 2. The red curve shows the interpolation polynomial $f_1(x)$ and the blue curve shows $f_2(x)$.

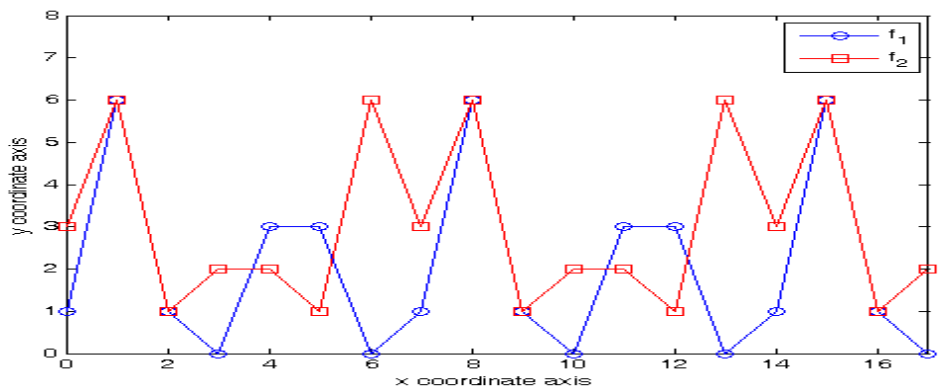

Figure 2. The curves for Example 2.